# An explicit solution to the spinning ring problem

Aradhya Jain

**Abstract**

A ring may be regarded as a torus with r << R, where R is the major radius and r is the minor radius. When such a ring is placed on a rough rod and released with some angular velocity, it may continue to vertically spin around the rod for some time instead of falling down immediately. It was observed that two different kinds of motion for the rod exist, which are referred to as single point and double point contact motion, based on the number of contact points of the ring that are in contact with the rod. Single point contact motion was observed for rings and double point contact motion was observed in the case of a washer. We investigated the characteristics of the single point contact motion. An explanation is provided for the single point contact motion of the ring and an analysis of the forces on the ring is made. We observe a hyperbolic decay in the angular velocity of the ring. An explicit solution is determined for the single point contact motion of the ring and the obtained solutions match the observed motion of the ring.

**Introduction:**

A ring/washer spinning on a rod may not fall when released because of the frictional forces acting on the body. Two different types of motion were observed in such a scenario based on the major radius of the ring/inner radius of the washer. The first is single point contact motion which is where the ring/washer touches the rod at only a single point. The second is double point contact motion which is where the ring/washer touches the rod at two points. Double point contact motion has already been investigated[1] and thus we will focus the scope of this paper on only single point contact motion. The frictional force is equal to the normal force times the coefficient of friction, where the normal force for such a body is equal to the centripetal force, $m\omega^2 r$. The diagram shows the free body diagram for the washer in both cases. The normal and centripetal forces are not displayed as they cancel each other out in all situations.

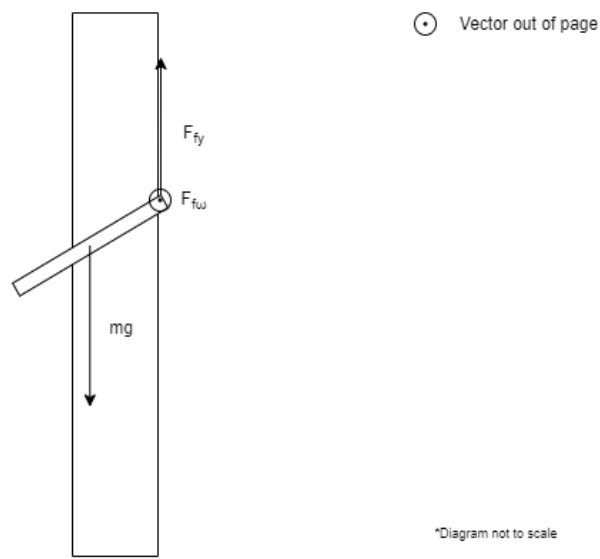

*Figure 1. Free body diagram[2] of Single point contact motion.*

As can be seen from the diagram, the frictional force that acts upon the ring may be resolved into two components - one in the theta direction and one in the y direction. The friction is proportional to the normal force which is equal to the centripetal force[3]. The coefficient of friction in the theta direction is the coefficient of rolling friction between the two materials whereas the friction in the y direction is the coefficient of static friction between the two materials. Let us define both of these values as $\mu_{rol}$ and $\mu$. On the basis of the force setup, we can create the following differential equation which describes the angular motion of the body. In the following, r is the *inner* radius of the torus, which is equal to the difference between the major and minor radii.

$$I\dot{\omega} = \vec{F} \times \vec{r} \Rightarrow mr^2\dot{\omega} = -mr^2\omega^2\mu_{rol} \tag{1.1}$$

The negative sign on the right hand side is present because friction is a resistive force.

This equation can be simplified to the following:

$$\frac{d\omega}{dt} = -\omega^2\mu_{rol} \tag{1.2}$$

Separating the variables and simplifying gives

$$\frac{1}{\omega\mu_{rol}} = t + c \tag{1.3}$$

At t = 0, $\omega = \omega_0$,

Substituting and simplifying further gives:

$$\boldsymbol{\omega(t) = \frac{1}{t\mu_{rol} + \frac{1}{\omega_0}}} \tag{1.4}$$

Intuitively, the following differential equation can be set up on the y axis.

$$\ddot{y} = g - r\mu\omega^2(t) \tag{2.1}$$

However, since the friction can never overcome a force, the following correction must be made to our formula for the differential equation.
This gives:

$$\ddot{y} = \frac{g - r\mu\omega^2(t) + |g - r\mu\omega^2(t)|}{2} \tag{2.2}$$

We can substitute the obtained function for ω(t) into the equation to get an equation that depends explicitly on time. Simplifying, we get:

$$\frac{d^2y}{dt^2} = \frac{g - \frac{r\mu\omega_0^2}{(\mu\omega_0 t + 1)^2} + \left|g - \frac{r\mu\omega_0^2}{(\mu\omega_0 t + 1)^2}\right|}{2} \qquad (2.3)$$

Integrating both sides gives us the following formula:

$$\frac{dy}{dt} = \frac{1}{2}\left(\left(gt + \frac{r\omega_0}{\mu\omega_0 t + 1}\right) + \left(sgn\left(g - \frac{r\mu\omega_0^2}{(\mu\omega_0 t + 1)^2}\right)\left(gt + \frac{r\omega_0}{\mu\omega_0 t + 1}\right)\right)\right) + c \qquad (2.4)$$

at t = 0, v(t) = 0 which gives:

$$\frac{1}{2}sgn(g - r\mu\omega_0^2)(r\omega_0) + \frac{r\omega_0}{2} + c = 0 \qquad (2.5)$$

at t = 0, all test cases involved $g < r\mu\omega_0^2$ which gives c = 0.

$$\frac{dy}{dt} = \frac{1}{2}\left(\left(gt + \frac{r\omega_0}{\mu\omega_0 t + 1}\right) + \left(sgn\left(g - \frac{r\mu\omega_0^2}{(\mu\omega_0 t + 1)^2}\right)\left(gt + \frac{r\omega_0}{\mu\omega_0 t + 1}\right)\right)\right) \qquad (2.6)$$

Taking the integral again under the same assumption as the previous integral gives the constant of integration as 0 and

$$y(t) = \frac{1}{2}\left(\left(\frac{1}{2}gt^2 + \frac{r}{\mu}ln(\mu\omega_0 t + 1)\right) + sgn\left(g - \frac{r\mu\omega_0^2}{(\mu\omega_0 t + 1)^2}\right)\left(\frac{1}{2}gt^2 + \frac{r}{\mu}ln(\mu\omega_0 t + 1)\right)\right) \qquad (2.7)$$

In order to test our results, readings were taken for ω(t).

A vertical, smooth iron rod was fixed onto a base. A slow-motion camera recording at 240 fps was setup to record the entire system. Rings of different inner radii were spun and the angular velocities calculated from the camera recordings.

Regression analysis[4] was performed on the data collected. A graph of the form $\frac{1}{at+b}$(reciprocal linear) was fitted, where a = μ and b = $\frac{1}{\omega_0}$ and the following results were obtained. In the data shown below, each ring is assigned a number called the Ring Ratio which is the ratio of the inner radius of the ring to the radius of the rod.

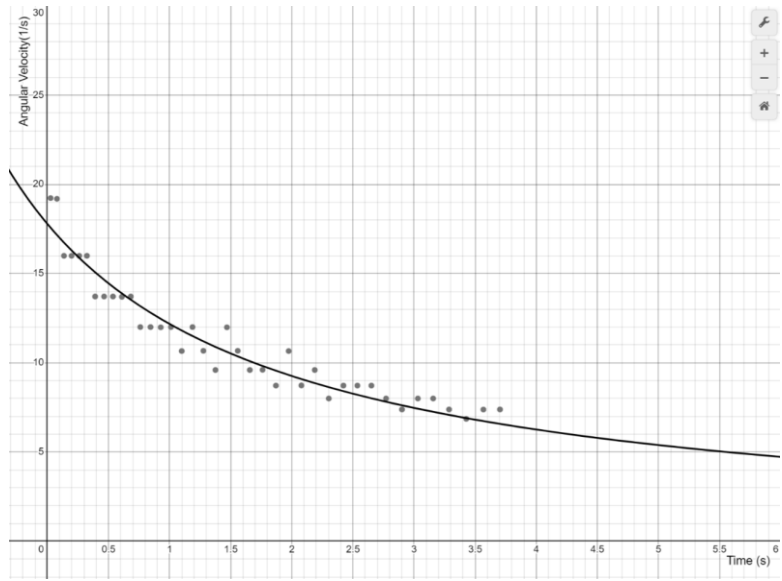

*Figure 2.1 – Reciprocal Linear Regression Analysis for a Ring with ratio 6.21*

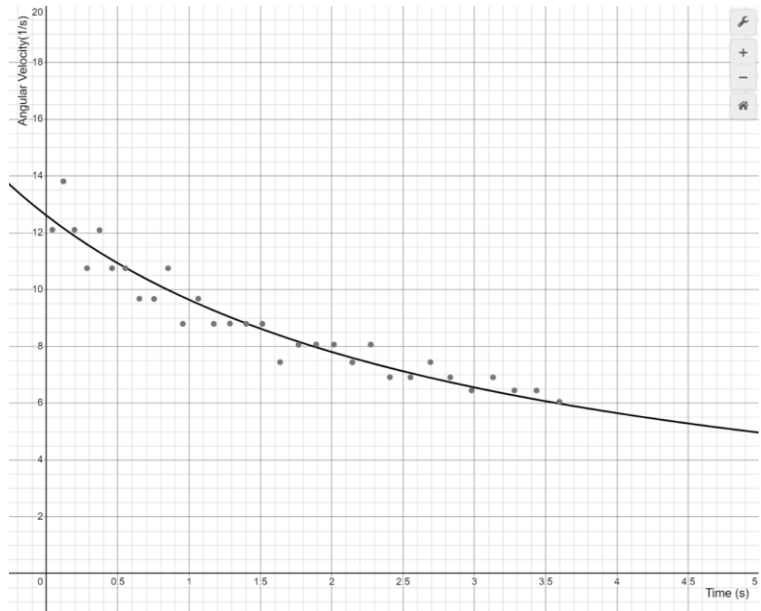

*Figure 2.2 – Reciprocal Linear Regression Analysis for a Ring with ratio 7.38*

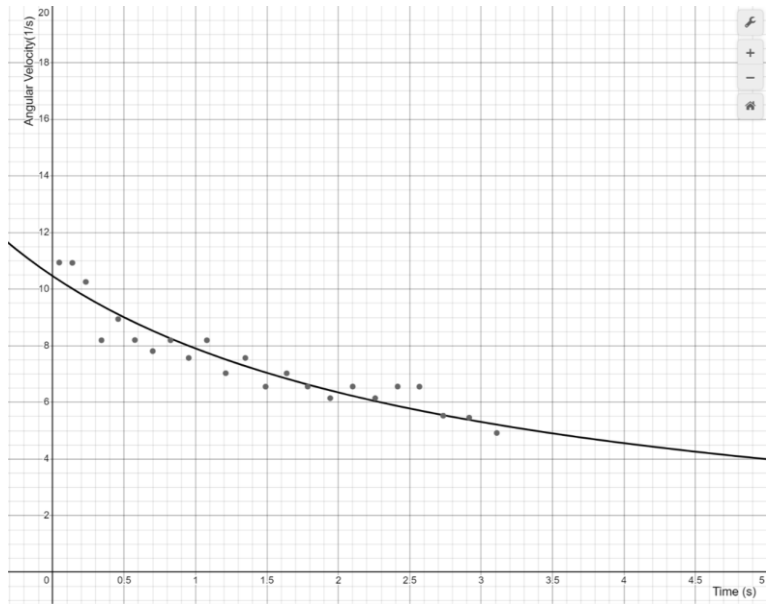
*Figure 2.3 – Reciprocal Linear Regression analysis for a Ring with ratio 9.35*

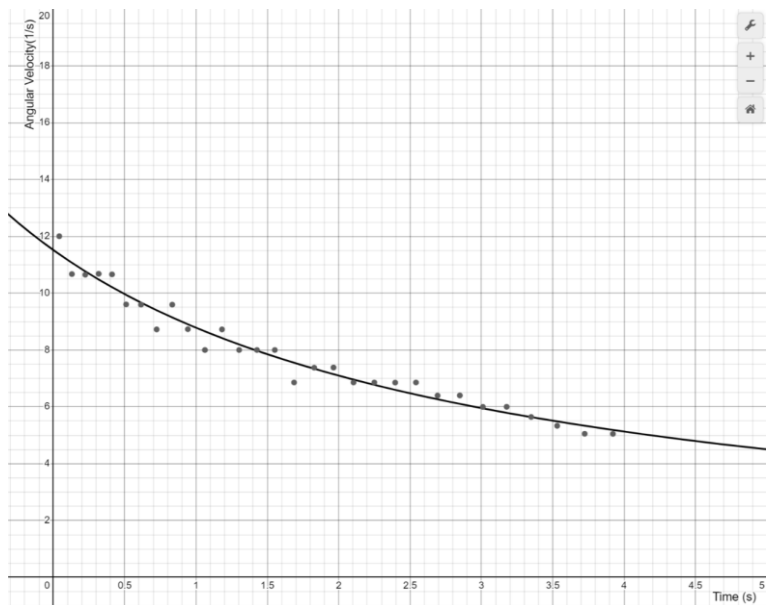
*Figure 2.4 – Reciprocal Linear Regression analysis for a Ring with ratio 10.21*

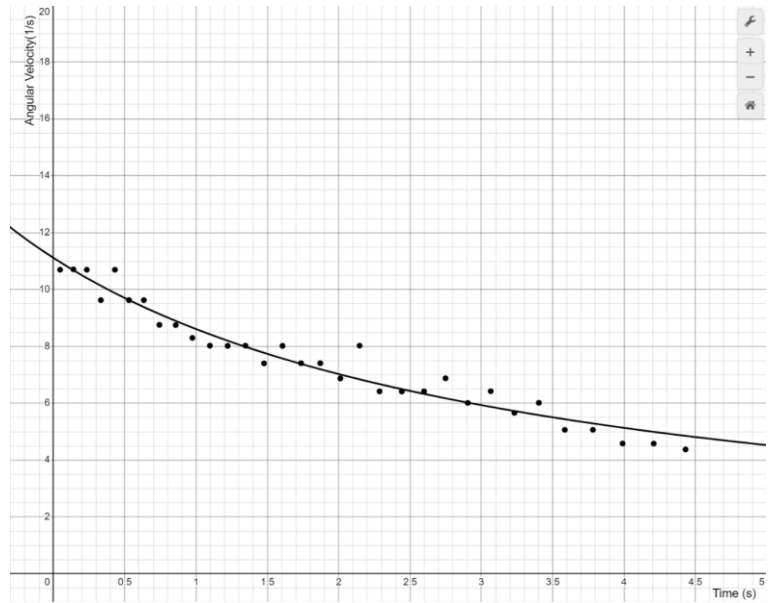

*Figure 2.5 – Reciprocal Linear Regression analysis for a Ring with ratio 10.21*

The table below shows the regressed values of μ for each ring, as well as the two materials which make up the surfaces.

| Ring Ratio | μ | Material 1 | Material 2 |
| --- | --- | --- | --- |
| 6.21 | 0.0258687 | Steel | Iron |
| 7.38 | 0.0243967 | Steel | Iron |
| 9.35 | 0.0309157 | Steel | Iron |
| 10.21 | 0.0270257 | Steel | Iron |
| 11.92 | 0.0261766 | Steel | Iron |

Online databases show that the value of μ is approximately around 0.02-0.03[5][6] based on the smoothness of the surfaces. As can be seen from the obtained data, the obtained results match with the online values of μ.

It can also be seen that the $r^2$ value for our model is high which means that the model explains much of the observed statistical deviance from the data.

Accurate data collection was not possible for the vertical position of the rod, as all camera setups resulted in data collected with high parallax error. However, it is expected that the calculated explicit solution will be followed in a similar manner to the angular velocity.

**Conclusion:**
We have determined the explicit solutions of the motion of the spinning ring on the vertical rod. The collected data shows that the angular velocity of the ring follows the expected hyperbolic decay. It is expected that the vertical motion of the ring also follows the explicit solutions calculated.

**Data Availability:**
Data will be made available from the author AJ upon reasonable request.